# Machine Learning for Yield Curve Feature Extraction: Application to Illiquid Corporate Bonds (Preliminary Draft)*

Greg Kirczenow[1], Ali Fathi[2] and Matt Davison[3]

[1]greg.kirczenow@gmail.com
[2]afathiba@uwo.ca
[3]mdavison@uwo.ca

**Abstract**

This paper studies the application of machine learning in extracting the market implied features from historical risk neutral corporate bond yields. We consider the example of a hypothetical illiquid fixed income market. After choosing a surrogate liquid market, we apply the Denoising Autoencoder algorithm from the field of computer vision and pattern recognition to learn the features of the missing yield parameters from the historically implied data of the instruments traded in the chosen liquid market. The results of the trained machine learning algorithm are compared with the outputs of a point-in-time 2 dimensional interpolation algorithm known as the Thin Plate Spline. Finally, the performances of the two algorithms are compared.

## 1. Introduction

In many fixed income markets it can be difficult to find bond yields for all rating and tenor pairs. This can arise in situations where market liquidity is insufficient to facilitate price discovery, either due to noisy or incomplete data. The central idea of the present paper is to design an unsupervised machine that learns the features of the corporate bonds yield curves by observing a sufficient number of historical examples in a liquid market, and then uses the learned shapes to fill in the missing yields in the illiquid market. Roughly speaking, the machine is a tool for interpolation/extrapolation, however, it also incorporates its memory of typical yield curve shapes.

This paper is highly influenced by algorithms that have been successfully employed for pattern recognition and computer vision. Specifically, our algorithm of choice is the denoising autoencoder (DAE) which is popular in image reconstruction problems and also training deep neural networks (see [(1)] and [(9)] for background). More concretely, we consider the example of a hypothetical illiquid fixed income market. Because the market is illiquid, reliable implied yields are available for only a limited set of traded ratings/tenors. After choosing a surrogate liquid market, we apply the DAE algorithm to learn the features of the missing yields from the historically implied data of the instruments traded in the chosen liquid market.

The results of the trained machine learning algorithm are compared with the outputs of a point-in-time 2 dimensional interpolation algorithm known as the Thin Plate Spline (TPS). While the TPS algorithm is a fairly flexible algorithm, in our view, it is an inferior tool since it only interpolate/extrapolates the rating/tenor grid on the spot without any insight of how the shape of the yield curve for a given rating should look. As a result, it can produce comparatively poor estimates in cases where key areas in the surface are missing.

The paper is organized as follows. In Section 2, the setup of the problem is explained. In Section 3, a high level description of the DAE algorithm and various network architectures for DAEs is given. For the sake of completeness, the details of the TPS algorithm are included. In Section 4, the details of the used data and algorithms implementation is presented. Finally, the results of the experiments are summarized in Section 5.

## 2. Problem Definition

The problem setup is fairly straight forward. Consider a fixed income trader who is supposed to provide quotes on prices for a variety of illiquid corporate/sovereign bonds for different ratings and tenors in a specific market. The trader only has access to bond yields for a few anchor points on the rating/tenor grid and she is tasked to complete the rating/tenor matrix based on the given points.

This is a problem of interpolation/extrapolation under market no-arbitrage constraints. These constraints are embodied collectively as the features present in the shape of the curves. For instance, in a normal market environment, the yield curves for each rating are upward sloping, with longer term interest rates being higher than shorter term. The curve might exhibit other features such as humps and mixes etc., based on the supply and demand for the instruments. In typical situations for our problem some of the yields at the short end of the curves for some of the ratings are known. There exists numerous approaches for interpola-

---

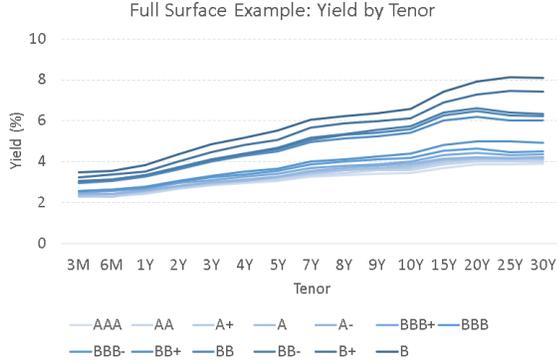

Figure 1: A typical yield curve structure (source: Bloomberg)

tion of the yield curve along the tenor coordinate (see [(2)] for a survey of the methods). However, the situation we are interested in consists of filling in a sparse matrix grid along both the rating and the tenor coordinates, therefore a two dimensional generalization of the above referred methods is needed.

Thin Plates Splines (TPS), [(3),(6), (10)] are natural analogues of the 1-D smoothing splines to 2-D, and are effective tools for filling in a 2-D grid and transforming it to a surface while the oscillatory behavior of the interpolating surface can be controlled (see Section 3.3 below). As noted in [(2)], spline interpolators do not necessarily produce the empirically expected features in the yield curve. This is also true for the 2-D TPS algorithm. Therefore, the raw outputs of the TPS algorithm most undergo an additional modification step in order to produce the desired curve features (such as monotonicity etc.).

Another subtlety involved in applying the ordinary interpolating algorithms to our case of sparse rating/tenor matrix is the assignment of a numerical coordinate vector to the rating dimension. In other words, in order to implement a 2-D interpolator such as TPS, an extra non-canonical criterion must be set in order to convert the ordinal values of the ratings to a quantitative coordinate vector.

## 3. Theoretical Background

In this section we provide a brief presentation of the theoretical background to understand the algorithms implemented in this paper. We start by defining the class of Autoencoder algorithms. Next we specify the sub class of this family of algorithms called Denoising Autoencoders. We also describe both the Fully Connected Neural Network (FCNN) and Convolutional Neural Network (CNN) architectures for these algorithms. Finally we introduce the Thin Plate Spline (TPS) algorithm which is used as the base line model in this paper. We refer the reader to [(1)] for details of NNs, CNNs and Autoencoders. The reader can consult [(3)] or [(10)] for more details on the TPS algorithm.

### 3.1 Autoencoders

In this section, we follow the exposition in [(1)] and [(9)] closely, where the reader may find the basics of neural networks. Autoencoders can be described as a class of unsupervised learning algorithms which are designed to transfer the inputs to the output through some (non-linear) reproducing procedure. They are called unsupervised since the algorithm only uses the inputs $X$ for learning. An autoencoder transforms an input vector $x \in \mathbb{R}^d$ to a hidden representation $h(x) \in \mathbb{R}^{d'}$ via a mapping $h(x) = g(Wx + b)$. $W$ is a $d' \times d$ weight matrix and $b$ is called a bias vector. The mapping $g(.)$ is also called the *encoder* part of the autoencoder.

The *decoder* part of the algorithm then maps the obtained hidden representation back to a reconstructed vector $z \in \mathbb{R}^d$ via a mapping $z = k(W^*h(x) + c)$.

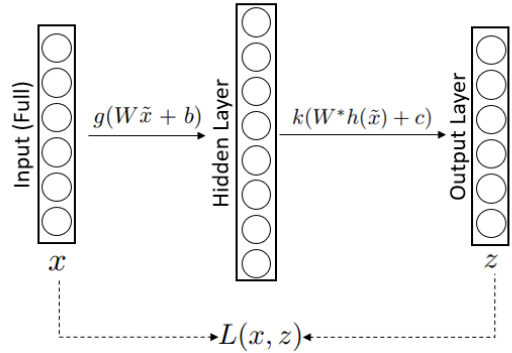

Figure 2: A general autoencoder architecture.

Therefore, observation $x_i$ is first mapped to a corresponding $h(x_i)$ and then to a reconstruction $z_i$. If we define $\theta = \{W, b\}$ and $\theta' = \{W^*, c\}$, the optimal weights are found by solving,

$$argmin_{\{\theta, \theta'\}} \frac{1}{N} \sum_{i=1}^{N} L(x_i, z_i). \qquad (1)$$

Here, $L(x, z)$ is a chosen *loss function*. A principal recipe in statistical learning for density estimation [(8)] can guide us in choosing the cost function for training the autoencoder. In general, one starts with choice of a joint distribution of the data $p(x|\theta)$, where $\theta$ is the vector of distribution parameters. Next, one defines relationship between and $\theta$ and $h(x)$ and sets up a loss function $L(k \circ g(x)) = -\log p(x, \theta)$, where $k \circ g(x) = z$ is the autoencoder functional form. For instance, the choice of square error loss $L(x, z) = ||x - z||^2$ is equivalent to the choice of a Gaussian distribution with mean $\theta$ and the identity covariance matrix for the data,

$$p(x, \theta) = \frac{1}{2\pi^{N/2}} exp(-\frac{1}{2}\sum_{i=1}^{N}(x_i - \theta_i)^2), \quad (2)$$

where $\theta = c + W^*h(x)$ (see [(1)]).

### 3.2 Denoising Autoencoders

The basic idea of an autoencoder as explained above is to minimize an expected loss function $\mathbb{E}[L(x, k \circ g(x)]$. Here, the loss function penalizes the *dissimilarity* of the $k \circ g(x)$ from $x$. This may drive the functional form to be merely an identity mapping $k \circ g(x) = x$ if the algorithm architecture allows for it. In other words if, the hidden layer $h(x)$ of the autoencoder is wider than input vector (overcomplete), the algorithm would just copy the inputs to the hidden layer and hence not learn meaningful features of the inputs.

To avoid this situation, a denoising autoencoder (DAE) can be trained. (This section follows the exposition in [(9)] closely; see [(1)] and references therein for more details.) The idea is to minimize the expected loss $\mathbb{E}[L(x, k \circ g(\tilde{x})]$ where $\tilde{x}$ is a *noisy* version of $x$. The DAE must therefore learn enough features in order to salvage the original data from the corrupted version rather than simply copying their input.

The process starts by first injecting noise to the initial input $x$ to obtain a partially corrupted copy $\tilde{x}$ via a stochastic procedure (so, $\tilde{x} \sim q_D(\tilde{x}|x)$. In [(9)] the following corruption process is implemented: for each input $x$, a fixed proportion $\nu$ (fixed in advance) of the coordinates are randomly set equal to 0 while the others are left untouched.* The corrupted copy of the input $\tilde{x}$ is then fed into a regular autoencoder (see Figure.1). It is important to notice that the parameters of the model are trained to minimize the average reconstruction error $\mathbb{E}[L(x, z)]$ not $\mathbb{E}[L(\tilde{x}, z)]$, namely to obtain $z$ as close as possible to the uncorrupted input $x$. Here, the output $z$ is related deterministically to $\tilde{x}$ rather than $x$.

### 3.3 Architectures for Autoencoders

The encoder and decoder parts of an autoencoder are feedforward neural networks with variable architectures (see [(9)] for definitions and details).

The Figure.2 above demonstrates the fully connected neural network (FCNN) for the DAE in this paper (see Section 4.2). It is seen in the figure that the input filtered through the encoder to produce the code. The decoder then reconstructs the output based on the code. The decoder architecture depicted above in Figure.2 is the mirror image of the encoder, this is not a requirement but a popular choice of architecture. The only requirement is the dimensions of the input space and the output space must be identical.

---

*An alternative way is to use Gaussian additive noise whose variance is set via hyper-parameter tuning (see [(1)]

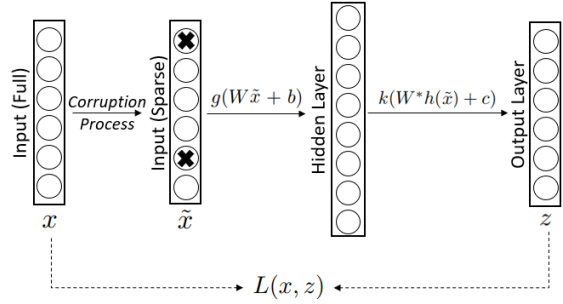

Figure 3: An autoencoder corrupts and then reconstructs the input.

As emphasized in [(5)], FCNN architectures for autoencoders and DAEs both ignore the 2D structure of an image (2D structure of the rating/tenor surface in our case). Hence, the most successful models in image pattern recognition capture the localized features that repeat themselves in the input (see [(5)] and references therein). These algorithms employ the CNN architectures for both the decoder and encoder parts of autoencoders (see Figure.4 and [(1)] for details on CNNs).

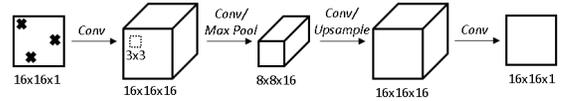

Figure 4: The optimal CNN network trained on historical rate surfaces.

### 3.4 Base Line Algorithm, Thin Plate Spline

Thin Plate Splines (TPS) are 2D generalizations of 1D cubic splines [see (3), (6), (10)]. TPS is an interpolation algorithm for the grid points $\{(X^i, y^i)\}_{i=1}^{m}$ where $X^i = (x_1^i, x_2^i) \in \mathbb{R}^2$. The spline surface $f(x_1, x_2)$ is constructed such that it passes the grid points as closely as possible while exhibiting a controlled amount of oscillation. This Bias/Variance optimal surface is obtained by minimizing the following action functional over an appropriate function space (see (10)).

$$S[f] = \sum_i |f(X_i) - y_i|^2 + \lambda \int |H(f)|^2 dx. \quad (3)$$

Here, $|H(f)|^2$ denotes the sum of square entries of the Hessian matrix of $f$ and $\lambda \in \mathbb{R}^+$ is a regularization parameter. The first term is a measure of fitting error and the integral in the second term measures the oscillating behaviour of the interpolation surface.

The TPS solution is the minimizer of the above defined action. Applying the Euler-Lagrange recipe for finding the minimum of the functional in (3) one obtains a 4th order PDE with certain orthogonality conditions on the space of solutions [see (3), (6)]. The solution in the 3D euclidean space (2D TPS) is given by the following closed form formula,

$$f(x) = \sum_{i=1}^{m} a_i G(X, X_i) + (b_0 + b_1 x_1 + b_2 x_2). \quad (4)$$

In practice, in the 3D case one can set $G(X, s) = u(|X-s|)$, where $u(x) = x^2 \log x$. Using the constraints $f(X_i) = y_i$, one obtains the following linear system,

$$Y = Ma + Nb, \quad (5)$$

where $M$ is an $m \times m$ matrix with the entries $M_{ij} = G(X_i, X_j)$ and $N$ is an $m \times 3$ matrix with the rows $[1 \ X_i^T]$. The system is subject to the orthogonality condition [see (6)] $N^T a = 0$. It turns out that the matrix $M$ is non-singular and the system (5) can be solved by,

$$b = (N^T M^{-1} N)^{-1} N^T M^{-1} Y, \quad (6)$$

$$a = M^{-1}(Y - Nb). \quad (7)$$

The hyper-parameter $\lambda$ must be set in advance. In practice, $\lambda$ is chosen through a hyper-parameter tuning process such as $k-$fold cross validation (see [(3)] and [(7)]).

## 4. Experimental Results

### 4.1 Data Description

The data are collected from Bloomberg (under the University of Western Ontario licence) and consist of mid-yields of corporate industrial bond indices. There are 13 indices (AAA, AA, A+, A, A-, BBB+, BBB, BBB-, BB+, BB, BB-, B+, B), and each index provides yields at 15 tenors (3m, 6m, 1y, 2y, 3y, 4y, 5y, 7y, 8y, 9y, 10y, 15y, 20y, 25y, 30y), resulting in a 2D surface with 195 points for each observation date. The yields are computed by Bloomberg from constituent bonds and are taken as given. The data are daily and range from Jan 29, 2018 to April 27, 2018 (63 observations).

Yields at 20, 25 and 30 years are unavailable for the BB+, BB, BB-, B+ and B indices. These missing points are populated by finding the spread between the missing tenor and the 15 year tenor for generic double and single B industrial corporate indices, and assuming that the same spread applies to the corresponding more granular rated indices.

It is observed that the yields are weakly monotonic in rating, with the exception of B+ and B indices at the long end of the curve for a single observation. Yields are also monotonic in tenor over the period observed, except in the very long end of the curve for some indices.

### 4.2 Training the Autoencoder

Training and test data for the neural networks are constructed as follows. First, the yields are scaled such that the maximum yield is 1. This aligns the model inputs and outputs with the range of the sigmoid activation function. Then, 10 percent of the observations are held out for testing. Since the purpose of the algorithm is to reconstruct a rate surface from known inputs, rather than predict a rate surface out of time, the test set is selected at random from the data.

Finally, to provide the algorithm with additional information for training, each observation from both the training and test sets is repeated 10 times and subjected to a randomized corruption procedure. Following ((9)), a fixed number of elements from each example are chosen at random and their value is forced to 0, creating sparse rate surfaces. Each neural network is then trained to reconstruct complete rate surfaces using the sparse surfaces as inputs.

The neural networks are implemented in Keras with TensorFlow backend. Adam optimization with standard parameter values and mean square error loss is used for both networks. The Python package `Hyperas`, with built-in Tree-of-Parzen-Estimators (TPE) algorithm ((11)), is used to conduct hyperparameter optimization for the learning rate, decay rate, batch size, number of nodes in the FCNN hidden layer, and number of CNN layers and filters. The Rectified Linear Unit (ReLU) is used as the activation function for hidden layers, and the sigmoid as the activation function for the output layer. Batch normalization is used after each hidden layer activation in both networks.

The FCNN uses a single overcomplete hidden layer (see Figure.2). Implicit in the structure is the idea that the network must learn relationships between each pair of ratings in the surface, no matter the distance between them in tenor/rating space. However, economic intuition suggests that adjacent yields on the surface will be strongly correlated. The CNN captures this feature in its network architecture, which consists of several layers of convolutions with 3x3 filters and pooling, followed by convolutions and up-sampling to restore the rate surface dimensions. As such, the CNN has considerably fewer parameters than the FCNN.

### 4.3 Fitting the Thin Plate Spline

The `Tps` functionality in the $R$-package `fields v9.6` was used in order to fit the TPS surfaces. We refer the reader to [(7)] for the details and set up of the functionality. The TPS algorithm was fitted to each of the 70 sparse matrices in the test set and then full matrix was predicted using the calibrated tps parameters. The results were then compared with actual full test matrices to derive the relevant error metrics (see Section 4.4 below).

### 4.4 Performance Testing

Average performance on the 70 test set examples is shown in the table below.

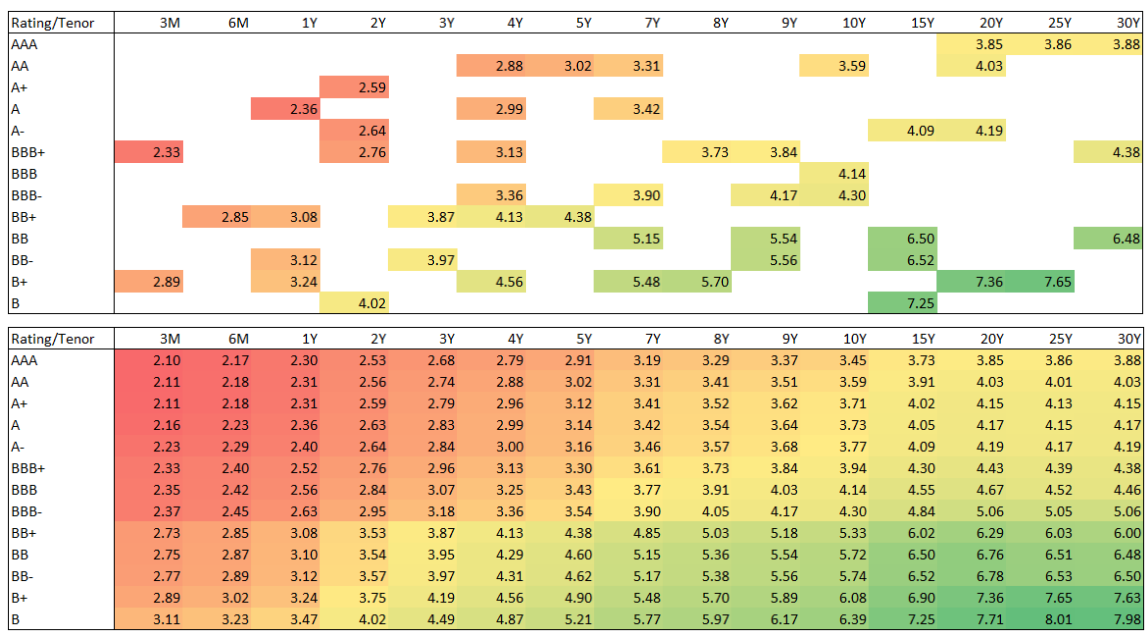

Figure 5: A sample full/ sparse yield matrix (Source: Bloomberg).

| Metric | TPS | FCNN | CNN |
|---|---|---|---|
| MAE (bps) | 11 | 8 | 10 |
| MAE (percent) | 3 | 2 | 3 |
| RMSE (bps) | 18 | 11 | 13 |
| RMSE (percent) | 4 | 3 | 3 |

Table 1: Test set performance $\nu = 0.75$

The Mean Absolute Error (MAE) and Root Mean Square Error (RMSE) are calculated over all points on the rate surface for all test set examples, and the values in percent are computed relative to the true observed yields.

For the Bloomberg data set, the two networks have similar performance. Recall that this time series is extremely short, with only 2 months of observations. The shape of the surface over this time period is rather stable.

The FCNN and CNN procedures were also run on a proprietary data set with a substantially longer history, including a variety of different surface shapes. For this data set, the CNN outperforms the FCNN, suggesting that it may have superior flexibility in learning a variety of tasks.

### 5. Summary of Results

This paper has demonstrated a novel financial application of well known algorithms in the image recognition literature. FCNN and CNN DAEs are shown to be capable of extracting features from liquid markets. Assuming that these same features are present in illiquid markets, the algorithms can be used to estimate missing information. This paper provides an example for corporate bonds, but similar approaches are likely to be fruitful in other areas such as equity volatility surfaces.

We note that the algorithm does not necessarily respect montonicity with respect to bond ratings in all cases. In the example we chose, about 10 percent of estimated yields violate this condition. While in principle this could be corrected by simply adjusting the loss function, we are more interested in finding a solution that allows the algorithm to discover this without prior knowledge of this feature.